# Magnetic nanowires as permanent magnet materials.

T. Maurer, F. Ott, G. Chaboussant,
Laboratoire Léon Brillouin CEA/CNRS UMR12 , Centre d'Etudes de Saclay, 91191 Gif sur Yvette, France

Y. Soumare, J.-Y. Piquemal,
ITODYS Université Paris 7- Denis Diderot, UMR CNRS 7086 2, place Jussieu 75251 cedex05 Paris, France

G. Viau,
Laboratoire de Physique et Chimie des Nano-Objets, INSA de Toulouse, UMR CNRS 5215,  135 av. de Rangueil 31077 Cedex 4 Toulouse, France

*Abstract :*
We present the fabrication of metallic magnetic nanowires using a low temperature chemical process. We show that pressed powders and magnetically oriented samples exhibit a very high coercivity (6.5 kOe at 140K, 4.8 kOe at 300K). We discuss the magnetic properties of these meta-materials and show that they have the suitable properties to realize "high-temperature magnets" competitive with AlNiCo or SmCo permanent magnets. They could also be used as recording media for high density magnetic recording.

In the 1960's, magnets made of elongated single domain (ESD) particles were commercialized under the name of Lodex [1-3]. However their performances were limited and they were superseded in the 70's by Rare-Earth (RE) magnets. In this letter, we revisit the subject and present an original way of synthesizing ESD particles which could be used to fabricate materials with a high remanence and a high coercivity. These ESD particles are nanowires made of 3d transition metals with potentially high Curie temperature. One interesting aspect is that they are produced using a low temperature chemical route enabling large scale fabrication. Moreover, the key advantage of these objects is their very large shape anisotropy which provides a high intrinsic coercivity to the material.

$Co_{80}Ni_{20}$ nanowires are synthesized by reduction in liquid polyol according to a procedure previously reported [4-5]. After synthesis, the black magnetic metal powder is collected by centrifugation and washed several times with ethanol. TEM observations show nanowires with a mean diameter $d_m$ of 6.8 nm and a mean length $L_m$ of 240 nm (Fig. 1). The standard deviation on the diameter distribution $\sigma_d$ is small ($\sigma_d/d_m = 10\%$) and the length distribution is broader with a standard deviation $\sigma_L = 55$ nm ($\sigma_L/L_m = 23\%$). X-ray diffraction patterns show mainly the metallic hcp phase and cobalt oxide CoO as a minor phase. High resolution electron microscopy shows very well crystallized wires with the crystallographic c-axis parallel to the wires. The overall composition of $Co_{80}Ni_{20}$ is evidenced by chemical analysis on powders.

The magnetic properties were characterized by standard magnetometry (SQUID and VSM). Dried powders were compressed. The density of the pressed powders was 4.4 g.cm$^{-3}$. A typical hysteresis cycle is presented on Figure 2. The sample exhibit a relatively high coercive field: $H_c = 286$ kA/m = 3.6 kOe. The saturation magnetization of the powder is in the range $8.2 \times 10^{-5} - 10.7 \times 10^{-5}$ T.m$^3$/kg (65 - 85 emu.g$^{-1}$). If normalized by the bulk density of the material, one obtains M = 581 – 758 kA/m (4πM = 7.3 - 9.4 kG). This is lower than the bulk values for Co$_{80}$Ni$_{20}$ alloys M = 1230 kA.m$^{-1}$ (4πM = 15.4 kG). This reduced magnetization is attributed to superficial oxidation of the nanowires (observed already with high resolution TEM). The Curie temperature T$_C$ of our nano-objects is expected to be relatively high since the magnetization is reduced by only 10% at 240°C (Fig 2). We estimate that T$_C$ is thus at least above 600°C.

Since the most interesting property of these nano-objects is their shape anisotropy (L$_m$/d$_m$>20), we tried to align the wires in order to increase the overall anisotropy of the material. In a first process, the particles were dispersed in liquid toluene. A magnetic field of 10 kOe was applied during the freezing of the solution (T$_f$ = 180 K). The magnetization was measured at 140K (see Figure 3). The square shape of the hysteresis loop suggests that the particles are well aligned in the solid toluene matrix. The characteristics of the sample are significantly improved: the coercive field is increased to 517 kA/m (6.5 kOe), compared to 286 kA/m for powder samples while the remanence is 0.974 M$_s$. Therefore, the nanowires alignment has significantly enhanced the anisotropy effects. This process however requires to work at low temperatures to freeze the particles solvent.

In order to produce solid samples at room temperature, a second process was devised. The particles were dispersed in a polymer solution (PMMA + toluene) and we let the material solidify under a magnetic field while the solvent was evaporating. The nanowires are expected to align along the magnetic field direction. We will refer to this direction as the easy axis while the perpendicular axis is referred to as the hard axis. Results obtained at T=300K are presented on Figure 4. Due to a partial alignment of the particles, the magnetic characteristics have been improved compared to the bulk samples. The remanence has increased to 0.85 M$_s$ for the nanowires aligned along the easy axis (compared to 0.65 M$_s$ for powder samples). Similarly, the coercive field has increased from 286 kA/m (3.6 kOe) to 381 kA/m (4.8 kOe).

The two different processes discussed above show that a significant coercivity increase can be achieved through nanowire alignment at either low temperature (140K) or room temperature. A direct critical comparison of these techniques should however consider the significant effect of temperature on the magnetocrystalline anisotropy [6]. Here, we are in the regime T<<T$_C$ and thus the shape anisotropy, which depends on the saturated magnetization M$_s$ and the demagnetising factor N like K ~ (1-N)M$_s^2$, is virtually temperature-independent.

To extract quantitative information about the ordering of the particles in the frozen solution and in the polymer matrix, we considered the Stoner-Wohlfahrt model [7]. This model is well suited since our nanowires are a limiting case of very elongated ellipsoids, almost monodispersed and with excellent cristallinity. The diameter of the wires is sufficiently small to ensure that we have single domain states in our wires [7,8] and we make the assumption that the wires do not interact one with the others. We consider three magnetic energies: the magnetostatic energy, the magneto-crystalline energy and the Zeeman energy. The free parameters of the model are the angular distribution of the wire directions which is assumed to be Gaussian and the magneto-crystalline anisotropy field. The magnetization measurement on Figure 3 has been modelled with a reasonable accuracy by considering a dispersion of the nanowires orientation of σ$_θ$ = 10° (HWHM) and by introducing a magneto-crystalline anisotropy field H$_{MC}$=3 kOe. The distribution of the wires orientations was chosen so as to fit the measured remanence. The magneto-crystalline anisotropy was chosen to fit the coercive field. This leads to the conclusion that the observed coercivity H$_C$=6.5 kOe is the sum of two equivalent contributions: a magnetocrystalline anisotropy H$_{MC}$=3 kOe (which is very close to the value for bulk hexagonal cobalt) and a shape anisotropy which contributes to H$_{shape}$=3.5 kG. The contribution of the shape anisotropy is rather disappointing since, for perfectly aligned nanowires, one would expect H$_{shape}$ = M$_s$/2 ≈ 7.5 kG. However, these results must be put in perspective with results obtained on arrays of nanowires grown by electrochemical routes [8] which are perfectly aligned but never exhibit coercivities higher than 3 kOe even for Fe where the theoretical value could be as high as 10 kOe. We think that these good results are due to the mono-crystalline quality of the wires. The surface defects are very limited especially compared to nanowires grown in nanoporous alumina templates for which surface defects are often imprinted into the metallic wires.

We argue that these composite nanomaterials exhibit very promising magnetic properties, suggesting that they could be used in the fabrication of permanent magnets. Firstly, their single crystal structure provides an extra magneto-crystalline anisotropy and the use of 3d transition metals provides high remanence and high $T_C$. Most important, through their strong shape anisotropy, they provide a large coercivity, almost temperature-independent far below $T_C$, so that these nanowires could potentially be used as high temperature permanent magnets as are AlNiCo magnets. However, we still have to assess the magneto-crystalline anisotropy contribution at high temperature. Finally, the low temperature synthesis process can be easily scaled to large volumes.

In order to fabricate permanent magnets, it is necessary to keep the nanostructuring of the material which means that it would not be possible to reach a packing density of 1. We estimate that a density about half the bulk value can be used while keeping the shape anisotropy properties. This would lead to a magnetization of about 7.5 kG for the bulk material. For such a material, the energy product would be of the order of $(BH)_{max}$ = 3.5kOe x 3.5kG ≈ 12 MGOe. Such energy products are not competitive with high performances NdFeB permanent magnets [9]. However, preliminary high temperature measurements show a good stability up to 250°C (Fig. 2), thus we think that these materials could be serious contenders for permanent magnet applications at high temperatures. As shown in Figure 5, they could fill a gap between RE magnets like SmCo [9] and AlNiCo magnets [10]. RE magnets have higher coercivities but they show stronger softening upon warming while AlNiCo magnets exhibit much lower coercivities in the whole temperature range of interest. In addition, the fabrication method is a low temperature chemical process which does not need any advanced metallurgical skills. Another potential application of these materials is their use in magnetic recording media. Their properties are already twice as good as existing materials used for magnetic tape recording [11-12].

**Acknowledgment**
We are grateful to J.B. Moussy for his help during the VSM measurements, M. Viret for the SEM pictures and F. Herbst for the TEM images.

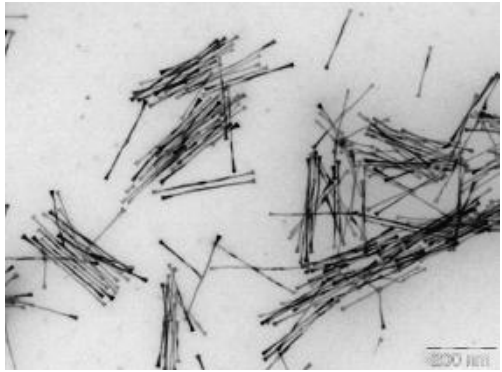

*Figure 1 : $Co_{80}Ni_{20}$ nanowires (TEM image).*

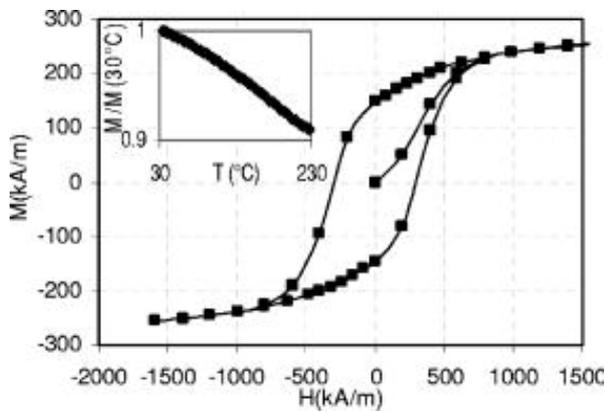

*Figure 2 : Magnetization (VSM) of a pressed powder sample (T=300K). (Inset) Variation of saturation magnetization with temperature.*

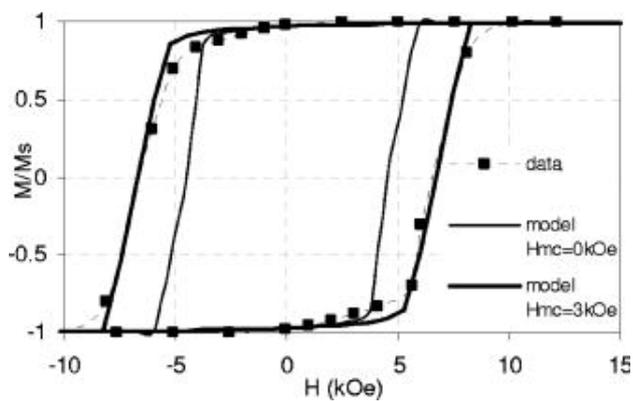

*Figure 3 : Magnetization (SQUID) of the nanowires dispersed in a frozen toluene matrix (solid line). Hysteresis cycles calculated using a Stoner-Wohlfahrt model with and without magneto-crystalline anisotropy (dashed lines).*

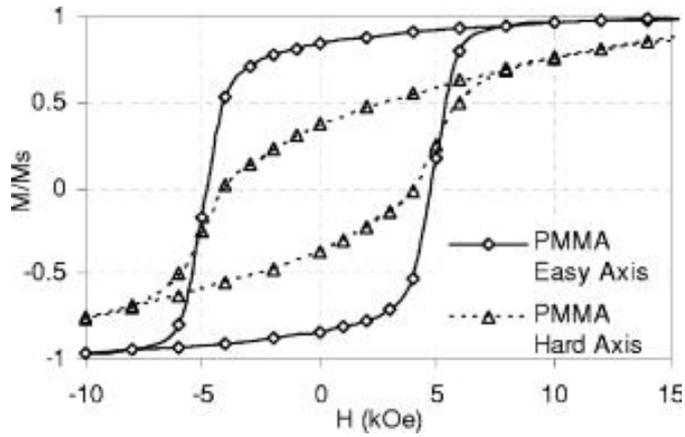

*Figure 4 : Magnetization of nanowires dispersed in PMMA (along the hard axis and the easy axis) at T=300K.*

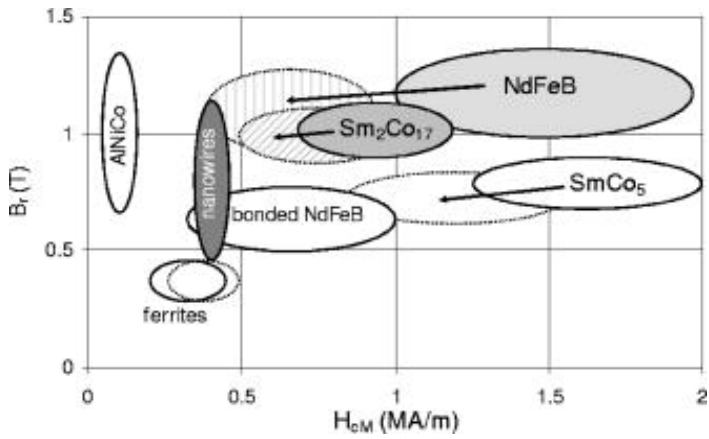

*Figure 5: Remanence and coercivity for the usual permanent magnets materials at 20°C. The values at 120°C are indicated as dotted/hashed surfaces. Nanowires, as AlNiCo or ferrites are not very temperature sensitive contrary to rare-earth magnets. Adapted from [13].*